\documentclass[twocolumn,floatfix,showpacs,prd,aps,tightenlines]{revtex4}
\usepackage{graphicx}
\usepackage{psfrag}
\usepackage{dcolumn}
\usepackage{bm}
\usepackage{amsmath}

\newcommand{\bea}{\begin{eqnarray}}
\newcommand{\eea}{\end{eqnarray}}
\newcommand{\beq}{\begin{equation}}
\newcommand{\eeq}{\end{equation}}
\newcommand{\KMS}{\rm km\,s^{-1}}

\begin{document}

\title{Foundations of multiple black hole evolutions}

\author{Carlos O. Lousto}
\affiliation{Center for Computational Relativity and Gravitation,
School of Mathematical Sciences,
Rochester Institute of Technology, 78 Lomb Memorial Drive, Rochester,
 New York 14623}

\author{Yosef Zlochower} 
\affiliation{Center for Computational Relativity and Gravitation,
School of Mathematical Sciences,
Rochester Institute of Technology, 78 Lomb Memorial Drive, Rochester,
 New York 14623}

\date{\today}

\begin{abstract} 

We present techniques for long-term, stable, and accurate evolutions of 
multiple-black-hole spacetimes using
the `moving puncture' approach with fourth- and eighth-order finite
difference stencils. We use these techniques to explore configurations
of three black holes in a hierarchical system consisting of a third
black hole approaching a quasi-circular black-hole binary, and find
that, depending on the size of the binary, the resulting encounter may
lead to a prompt merger of all three black holes, production of a
highly elliptical binary (with the third black hole remaining
unbound), or disruption of the binary (leading to three free black
holes).  We also analyze the classical Burrau three-body problem using
full numerical evolutions. In both cases, we find behaviors distinctly
different from Newtonian predictions, which has important implications
for N-body black-hole simulations.  For our simulations we use 
approximate analytic initial data.  We find that the eighth-order stencils
significantly reduce the numerical errors for our choice of grid
sizes, and that the approximate initial data produces the expected
waveforms for black-hole binaries with modest initial separations.

\end{abstract}

 \maketitle

\section{Introduction}
The recent dramatic breakthroughs in the numerical techniques to
evolve black-hole-binary
spacetimes~\cite{Pretorius:2005gq,Campanelli:2005dd,Baker:2005vv} has
led to rapid advancements in our understanding of black-hole physics.
Notable among these advancements are developments in mathematical
relativity, including systems of PDEs and gauge
choices~\cite{Gundlach:2006tw, vanMeter:2006vi}, the exploration of the
validity of
the cosmic censorship conjecture~\cite{Campanelli:2006uy,Sperhake:2007gu}, and
the application of isolated horizon
formulae~\cite{Dreyer:2002mx,Schnetter:2006yt,Campanelli:2006fg,
Campanelli:2006fy,Cook:2007wr,Krishnan:2007pu}.  There are many
exciting new results on recoil
velocities~\cite{Baker:2006vn,Sopuerta:2006wj,Gonzalez:2006md,
Sopuerta:2006et,Herrmann:2006cd,Herrmann:2007zz,Herrmann:2007ac,
Campanelli:2007ew,Koppitz:2007ev,Choi:2007eu,Gonzalez:2007hi,
Baker:2007gi,Campanelli:2007cga,Berti:2007fi,Tichy:2007hk,
Herrmann:2007ex,Brugmann:2007zj,Schnittman:2007ij,Krishnan:2007pu,
HolleyBockelmann:2007eh,Pollney:2007ss,Lousto:2007db}, post-Newtonian
(PN) and numerical waveform
comparisons~\cite{Baker:2006ha,Pan:2007nw,Husa:2007rh,Hannam:2007ik},
modeling of the remnant spin~\cite{Campanelli:2006uy,
Campanelli:2006fg,Campanelli:2006fy,Herrmann:2007ex,Rezzolla:2007xa,
Marronetti:2007wz,Rezzolla:2007rd}, new studies of eccentric black
hole binaries \cite{Pretorius:2007jn,Sperhake:2007gu,Hinder:2007qu}
and producing waveforms for matched
filtering~\cite{Vaishnav:2006ci,Baker:2006kr,
Ajith:2007qp,Vaishnav:2007nm,Buonanno:2007pf,Baker:2007fk,
Ajith:2007kx}.  In particular, the recent discovery of very large
merger recoil kicks for black-hole binaries with spins in the orbital
plane, which was originally inferred from the results
in~\cite{Campanelli:2007ew}, then observed in~\cite{Gonzalez:2007hi},
and determined to have a maximum value of $4000\ \KMS $
in~\cite{Campanelli:2007cga},  has had a great impact in the
astrophysical community, with several groups now seeking for
observational traces of such high speed holes as the byproduct of
galaxy collisions~\cite{Bonning:2007vt,HolleyBockelmann:2007eh}.

Three-body and four-body interactions are expected to be common in
globular clusters~\cite{Miller:2002pg,Gultekin:2003xd} and in
galactic cores hosting supermassive black holes (when
stellar-mass-black-hole-binary systems interact with the supermassive
black hole).  Hierarchical triplets of supermassive black holes might
also be formed in galactic nuclei undergoing sequential
mergers~\cite{Makino:1990zy,valtonen96}.  The gravitational wave
emission from such systems was recently estimated using post-Newtonian
techniques~\cite{Gultekin:2005fd}.  The recent discovery of a probable
triple quasar~\cite{Djorgovski:2007ka} (with estimated masses of
$50\times10^6 M_\odot$, $100\times10^6M_\odot$ and $500\times10^6
M_\odot$ and projected separations of between $30$ and $50$ kpcs)
indicates that hierarchical supermassive-three-black-hole systems are
possible.  Triple stars and black holes are much more common in
globular clusters~\cite{Gultekin:2003xd}, and galactic disks. The
closest star to the solar system, Alpha Centauri, is a triple system,
as is Polaris and HD 188753.

On the theoretical side, even in the Newtonian theory of gravity,
the three-body problem is much more complicated than
the two body one, and is generically chaotic.
In Ref.~\cite{Campanelli:2007ea} we established that our 
numerical formalism is able to handle the evolution of three fully
relativistic bodies.  In this paper we continue our quest to expand
our understanding of the multiple-black-hole problem by studying a
sequences of black-hole-binary---third-black-hole configurations,
with the third black hole intersecting the binary along the
axis of rotation,
to
explore the influence of a third black hole on the binary dynamics.
We also examine the classical Burrau
three-body problem and find that the dynamics can differ dramatically
from the Newtonian prediction, with important consequences for N-body
black-hole simulations.

This paper is organized as follows: in Sec.~\ref{sec:tech} we describe
the techniques used for multiple black-hole evolutions, including
eighth-order techniques and approximate initial data. In
Sec.~\ref{sec:res} we show the results from multiple black-hole
evolutions of three families of hierarchical configurations consisting of a
third black hole interacting with a black-hole binary along the
binary's axis of rotation, as well as the classical Burrau three-body
problem. We conclude our analysis in Sec.~\ref{sec:disc} where we
discuss the implications of our numerical results to black-hole
astrophysics and N-body simulations. We provide the techniques to
generate 2PN quasi-circular orbits of our hierarchical three-black-hole
systems in Appendix~\ref{ap:3-body-H}.

\section{Techniques} \label{sec:tech}
We evolve the three-black-hole-data-sets using the
{\sc LazEv}~\cite{Zlochower:2005bj} implementation
of the `moving puncture approach'~\cite{Campanelli:2005dd,Baker:2005vv}.
In our
version of the moving puncture approach~\cite{Campanelli:2005dd} we
replace the BSSN~\cite{Nakamura87,Shibata95, Baumgarte99} conformal
exponent $\phi$, which has logarithmic singularities at the punctures,
with the initially $C^4$ field $\chi = \exp(-4\phi)$.  This new
variable, along with the other BSSN variables, will remain finite
provided that one uses a suitable choice for the gauge.
We use the Carpet~\cite{Schnetter-etal-03b} driver to
provide a `moving boxes' style mesh refinement. In this approach
refined grids of fixed size are arranged about the coordinate centers
of each hole.
The Carpet code then moves these fine grids about the
computational domain by following the trajectories of the three black
holes.
We use {\sc AHFinderDirect}~\cite{Thornburg2003:AH-finding} to locate
apparent horizons.

We obtain accurate, convergent waveforms and horizon parameters by
evolving this system in conjunction with a modified 1+log lapse and a
modified Gamma-driver shift
condition~\cite{Alcubierre02a,Campanelli:2005dd}, and an initial lapse
$\alpha(t=0) = 2/(1+\psi_{BL}^{4})$ (where $\psi_{BL}$ is the
Brill-Lindquist conformal factor discussed below).
The lapse and shift are evolved with
$(\partial_t - \beta^i \partial_i) \alpha = - 2 \alpha K$, $\partial_t
\beta^a = B^a$, and $\partial_t B^a = 3/4 \partial_t \tilde \Gamma^a -
\eta B^a$.
These gauge conditions require careful treatment of $\chi$, the
inverse
of the three-metric conformal factor,
near the puncture in order for the system to remain
stable~\cite{Campanelli:2005dd,Campanelli:2006gf,Bruegmann:2006at}.
As was shown in Ref.~\cite{Gundlach:2006tw},
this choice of gauge leads to a strongly hyperbolic
evolution system provided that the shift does not become too large.
Unless otherwise noted, we use a standard choice of $\eta=6/M$ for the
simulations below.

\subsection{Approximate Initial Data}
We use the puncture approach~\cite{Brandt97b} to
compute initial data (See also~\cite{Beig94,Beig96} for similar
approaches).
  In this approach the 3-metric on the initial
slice has the form $\gamma_{a b} = (\psi_{BL} + u)^4 \delta_{a b}$,
where $\psi_{BL}$ is the Brill-Lindquist conformal factor,
$\delta_{ab}$ is the Euclidean metric, and $u$ is (at least) $C^2$ on
the punctures.  The Brill-Lindquist conformal factor is given by
$
\psi_{BL} = 1 + \sum_{i=1}^n m_i / (2 r_i),
$
where $n$ is the total number of `punctures', $m_i$ is the mass
parameter of puncture $i$ ($m_i$ is {\em not} the horizon mass
associated with puncture $i$), and $r_i$ is the coordinate distance to
puncture $i$. We solve for $u$ using the approximate solutions given
in Refs.~\cite{Gleiser:1997ng,Gleiser:1999hw,Laguna:2003sr,Dennison:2006nq},
with the addition of a cross-term
given below. We will use the
notation of~\cite{Laguna:2003sr} to write these approximate solutions.
For a spinning puncture, $u$ has the form (Eqs.~(4.23),(4.26),(4.27)
of~\cite{Laguna:2003sr}):
\begin{equation}
  \label{eq:approx_j}
  u_{J} (\ell, \mu_J) ={\cal J}^2 \left(u_{0}^J + u_{2}^J
 {\cal R}^2 P_2 (\mu_J)\right),
\end{equation}
where
\begin{subequations}
\begin{eqnarray}
  \label{eq:approx_jdef}
  40 u_{0}^J &=& \ell + \ell^2 +\ell^3 - 4 \ell^4 + 2 \ell^5,\\
  20 u_{2}^J &=& - \ell^5,
\end{eqnarray}
\end{subequations}
$\vec {\cal J} = 4 \vec J/m^2$, $m$ is the mass parameter,
$\ell = 1/(1+{\cal R})$, $\vec {\cal R} = 2 \vec r /m$,
   $\mu_J = \hat J\cdot \hat r$, and $P_2(x) = (3 x^2-1)/2$ is the
Legendre polynomial of degree 2.
For a boosted puncture, $u$ has the form (Eqs.~(4.44),(4.48),(4.49)
of~\cite{Laguna:2003sr}):
\begin{equation}
  \label{eq:approx_p}
  u_{P} (\ell, \mu_P) = {\cal P}^2(u_{0}^P + u_{2}^P P_2(\mu_P)),
\end{equation}
where
\begin{subequations}
\begin{eqnarray}
  \label{eq:approx_pdef}
  \frac{32}{5} u_{0}^P &=& \ell - 2 \ell^2 + 2 \ell^3 -
\ell^4+\frac{1}{5} \ell^5,\\
 80 {\cal R}\, u_{2}^P &=&  15 \ell +132 \ell^2 + 53 \ell^3 + 96 \ell^4+82
\ell^5+ \nonumber  \\
  && 84 \ell^5 /{\cal R} + 84 \ln(\ell)/{\cal R}^2, 
\end{eqnarray}
\end{subequations}
$\vec{\cal P} = 2 \vec P /m$ and $\mu_P = \hat P\cdot \hat r$.
If the puncture is both boosted and spinning then there is a cross-term
\begin{eqnarray}
  \label{eq:approx_c}
  u_c = (\vec{\cal P}\times\vec{\cal J})\cdot \vec {\cal R}
   (1+ 5 {\cal R} + 10 {\cal R}^2) \ell^5/80.
\end{eqnarray}
For a spinning boosted puncture
the  approximate solution has the form
\begin{eqnarray}
  \label{eq:approx_u}
  u &=& u_{P} + u_{J} + u_{c} + {\cal O}({\cal P}^4) +
  {\cal O}({\cal J}^4)  \nonumber \\
  &+& {\cal O} ({\cal P} {\cal J}^3) +
 {\cal O} ({\cal P}^2 {\cal J}^2) +  {\cal O} ({\cal P}^3 {\cal J}),
\end{eqnarray}
where the error terms linear in ${\cal P}$ and ${\cal J}$ only occur when
$\vec {\cal P}\times \vec {\cal J} \neq 0$.  We obtain
solutions for multiple punctures by superposition. The error in this
approximation scales as the inverse of the distance squared between
punctures.

In this approximation the total ADM mass, linear momentum, and
angular momentum are given by:
\begin{eqnarray}
M_{\rm ADM} &=& \sum_i m_i + \frac{2}{5} J_i^2/m_i^3 +
  \frac{5}{8} P_i^2/m_i,\label{eq:approx_MADM}\\
\vec P_{\rm ADM} &=& \sum_i \vec P_i,\\
\vec J_{\rm ADM} &=& \sum_i (\vec J_i + \vec r_i \times \vec P_i).
\end{eqnarray}
 
\begin{figure}
\begin{center}
\caption{A comparison of the $(\ell=2, m=2)$ modes of $\psi_4$ at
$r=40M$ for an equal-mass, non-spinning, quasi-circular binary of mass $1M$ and orbital frequency
$M\Omega = 0.035$, produced using the {\sc TwoPunctures} thorn (labeled
`Exact') and the approximate data (labeled `Approx'). In both cases we
used identical values for the initial data parameters (puncture mass,
puncture locations, puncture momenta). The approximate data has been
translated by $\delta t = 49.5M$ and multiplied by a constant phase
factor of $\exp(-3.8289\,i)$. The third plot (`Approx\_Renorm') shows
the waveform using the approximate data with the puncture masses
rescaled such that the ADM mass as given by Eq.~(\ref{eq:approx_MADM})
is 1M. This latter plot has been
translated by $\delta t=-62M$ and multiplied by a phase factor of
$\exp(4.5772\,i)$.  The inset shows an expanded view of the orbital
motion. Note the excellent agreement in the waveforms. }
\includegraphics[width=3.3in]{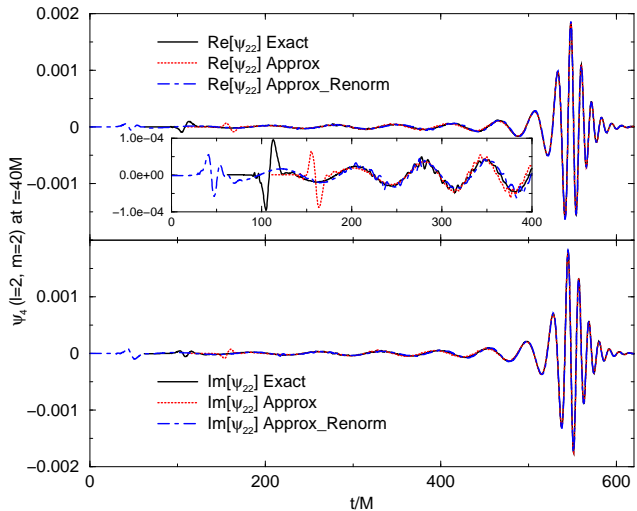}
\label{fig:data_compare}
\end{center}
\end{figure}

We test the reliability of this approximate initial data by solving
for equal-mass, non-spinning, quasi-circular binaries using both the {\sc
TwoPunctures}~\cite{Ansorg:2004ds} thorn (which is restricted to
problems involving two punctures) and
Eq.~(\ref{eq:approx_u}). We use the 3PN equations of motion to obtain
position and momentum parameters for two black holes in a binary with
unit mass and a specified orbital frequency. We then choose puncture
mass parameters (the same for both holes) such that the total ADM mass,
as calculated using the {\sc TwoPunctures} thorn, is $1M$.
For our test case we
choose a binary with orbital frequency $M\Omega = 0.035$ (which
performs at-least three orbits prior to merger) and evolve
using a relatively coarse resolution of $h=M/32$ (on the finest grid).
As shown in Fig.~\ref{fig:data_compare}, we find that the two
waveforms agree after the first half-cycle of orbital motion (we
corrected for a time translation and phase difference in the waveform
using the techniques of~\cite{Baker:2006vn,Baker:2007fb}).
The phase difference and time translation between the waveforms
produced by the two methods seems to be a result of the normalization
of the approximate initial data (i.e.\ the puncture mass parameters).
When we normalize the approximate
initial data such that the
total approximate ADM mass (i.e. Eq.~(\ref{eq:approx_MADM})) is 1M (it
is only exactly 1M for the {\sc TwoPunctures} data), then the binary
takes longer to inspiral and the phase difference and time translation
are in the opposite direction.  We find, based on a linear
interpolation of the phase difference and time translation versus
puncture mass, that there is a puncture mass choice where both the
phase difference and time translation are zero. The puncture mass that
gives an ADM mass of $1M$ for the exact initial data is $m_p =
0.4891M$, the puncture mass that gives and ADM mass of $1M$ for the
approximate data is $m_p = 0.4846M$, and the interpolated puncture mass
that gives phase (and time) agreement between the exact and
approximate waveforms is $m_p = 0.4871M$.

\subsection{Eighth-Order Finite Differencing}
We use both a new eighth-order spatial finite differencing algorithm
and the standard fourth-order finite differencing used in our previous
papers. Our eighth-order scheme extends the sixth-order scheme
described in Ref.~\cite{Husa:2007hp}. As in~\cite{Husa:2007hp}, we use
a fourth-order Runge-Kutta time integrator and a second-order in time,
fifth-order in space prolongation operator. Centered first spatial
derivatives have the form:
\begin{eqnarray}
\partial_x f_i &=& (3 f_{i-4} - 32 f_{i-3} + 168 f_{i-2}
      - 672 f_{i-1} \nonumber \\
      &-&\ 3 f_{i+4} + 32 f_{i+3}  -168 f_{i+2} + 672 f_{i+1})\nonumber \\
      &\,& / (840\,dx)\label{eq:dx_cen}
\end{eqnarray}
(we suppress the other two indices in
Eqs.~(\ref{eq:dx_cen})-(\ref{eq:dx2_cen})),
while for advection derivatives we adjust the center of the stencil by
one point. The downward pointing stencil has the form:
\begin{eqnarray}
  \partial_x f_i &=& (-3 f_{i-5} + 30 f_{i-4} - 140 f_{i-3}
    + 420 f_{i-2} - 1050 f_{i-1} \nonumber \\
    &\,&\ + 378 f_{i} +420 f_{i+1} - 60 f_{i+2}
    +5 f_{i+3})\nonumber \\
      &\,& / (840\,dx), \label{eq:dx_dn}
\end{eqnarray}
while the upward pointing stencil has the form:
\begin{eqnarray}
  \partial_x f_i &=& (+3 f_{i+5} - 30 f_{i+4} + 140 f_{i+3}
    - 420 f_{i+2} + 1050 f_{i+1} \nonumber \\
    &\,&\ - 378 f_{i} -420 f_{i-1} + 60 f_{i-2}
    -5 f_{i-3})\nonumber \\
    &\,& / (840\,dx). \label{eq:dx_up}
\end{eqnarray}
We use standard centered differencing for the second-spatial
derivatives.
The $\partial_{xx}$, $\partial_{yy}$, and $\partial_{zz}$ derivatives
have the form:
\begin{eqnarray}
\partial_{xx} f_i &=&\nonumber \\
 (&-&9 f_{i-4} + 128 f_{i-3} - 1008 f_{i-2} + 8064 f_{i-1} \nonumber \\
 &-&9 f_{i+4} + 128 f_{i+3} - 1008 f_{i+2} + 8064 f_{i+1} \nonumber \\
 &-&14350 f_{i} ) / (5040\ dx^2), \label{eq:dx2_cen}
\end{eqnarray}
while  the mixed spatial derivatives are obtained by applying
Eq.~(\ref{eq:dx_cen}) successively in the two direction, and
have the form:
\begin{eqnarray}
&\partial_{xy}& f_{i,j} = \nonumber\\
 &(& 9 f_{i-4,j-4} - 96 f_{i-4,j-3} + 504 f_{i-4,j-2} \nonumber\\
 &-& 2016 f_{i-4,j-1} + 2016 f_{i-4,j+1} - 504 f_{i-4,j+2} \nonumber\\
 &+& 96 f_{i-4,j+3} - 9 f_{i-4,j+4} - 96 f_{i-3,j-4} \nonumber\\
 &+& 1024 f_{i-3,j-3} - 5376 f_{i-3,j-2} + 21504 f_{i-3,j-1}\nonumber\\
 &-& 21504 f_{i-3,j+1} + 5376 f_{i-3,j+2} - 1024 f_{i-3,j+3} \nonumber\\
 &+& 96 f_{i-3,j+4} + 504 f_{i-2,j-4} - 5376 f_{i-2,j-3}\nonumber\\
 &+&28224 f_{i-2,j-2} - 112896 f_{i-2,j-1} + 112896 f_{i-2,j+1}\nonumber\\
 &-&28224 f_{i-2,j+2} + 5376 f_{i-2,j+3} - 504 f_{i-2,j+4}\nonumber\\
 &-& 2016 f_{i-1,j-4} + 21504 f_{i-1,j-3} - 112896 f_{i-1,j-2}\nonumber\\
 &+& 451584 f_{i-1,j-1} - 451584 f_{i-1,j+1} + 112896 f_{i-1,j+2}\nonumber\\
 &-& 21504 f_{i-1,j+3} + 2016 f_{i-1,j+4} + 2016 f_{i+1,j-4}\nonumber\\
 &-& 21504 f_{i+1,j-3} + 112896 f_{i+1,j-2} - 451584 f_{i+1,j-1}\nonumber\\
 &+& 451584 f_{i+1,j+1} - 112896 f_{i+1,j+2} + 21504 f_{i+1,j+3} \nonumber\\
 &-& 2016 f_{i+1,j+4} - 504 f_{i+2,j-4} + 5376 f_{i+2,j-3}\nonumber\\
 &-& 28224 f_{i+2,j-2} + 112896 f_{i+2,j-1} - 112896 f_{i+2,j+1}\nonumber\\
 &+& 28224 f_{i+2,j+2} - 5376 f_{i+2,j+3} + 504 f_{i+2,j+4}\nonumber\\
 &+&96 f_{i+3,j-4} - 1024 f_{i+3,j-3} + 5376 f_{i+3,j-2}\nonumber\\
 &-& 21504 f_{i+3,j-1} + 21504 f_{i+3,j+1} - 5376 f_{i+3,j+2}\nonumber\\
 &+& 1024 f_{i+3,j+3} - 96 f_{i+3,j+4} - 9 f_{i+4,j-4} \nonumber\\
 &+& 96 f_{i+4,j-3} - 504 f_{i+4,j-2} + 2016 f_{i+4,j-1} \nonumber\\
 &-&2016 f_{i+4,j+1} + 504 f_{i+4,j+2} - 96 f_{i+4,j+3}\nonumber\\
 &+& 9 f_{i+4,j+4} ) /(705600\,dx\,dy).\label{eq:dxdy}
\end{eqnarray}

We modify the stencils at the refinement (and outer) boundary zones
using the techniques proposed in
Refs.~\cite{Bruegmann:2006at,Husa:2007hp}.  The refinement boundary
points are not updated during timestep, but are updated by the
prolongation operation.  For the first through fourth points from the
boundary, we use standard second through eighth order techniques (as
described below), with the exception that we use centered derivatives
for the advection terms if the up(down)winded derivatives do not fit
on the grid.  The first points in from the boundary are updated using
standard second-order stencils, the second points using the standard
fourth-order scheme (See~\cite{Zlochower:2005bj}), the third points
using the standard sixth-order stencils (See Ref.~\cite{Husa:2007hp}),
and the fourth points using the proposed eighth-order scheme.  As in
Ref.~\cite{Husa:2007hp}, we found satisfactory results using 6 buffer
points. We use the standard fourth-order Kreiss-Oliger dissipation
operator.

\section{Results}\label{sec:res}

The initial data parameters for all new runs presented in this paper
are given in Tables~\ref{table:ID-1}--\ref{table:ID-4}.
For all of our three-black-hole runs we use a standard grid structure
consisting of 11 levels of refinement with outer boundaries at $640M$
and finest resolution of $h=M/80$. All runs were performed using
fourth-order techniques, except where otherwise noted.
\begin{table}
\caption{ 
Initial data parameters for configurations
with a third black hole intercepting a binary along the
$z$-axis. $(x_i,y_i,z_i)$ and $(p^x_i,
p^y_i,p_i^z)$ are the initial position and momentum of the puncture
$i$, $m^p_i$ is the puncture mass parameter, $m^H_i$ is the horizon
mass, $M \Omega$ is the binary's orbital frequency, and $D$ is the
binary's initial coordinate separation. Parameters not specified are zero.
Configurations are denoted by 3BHYXX, where ${\rm Y}=2, 3, 4, 5$ indicates
the momentum of the third black hole, with
  $p_3^z = -({\rm Y}-1) P_{0}$ (See Eq.~(\ref{eq:P0}).), 
and XX indicates the initial binary separation.
}
\begin{ruledtabular}
\begin{tabular}{lccc}
Config     &  3BH203      &  3BH205      &  3BH207      \\
\hline
$y_1/M$    &   3.6321068  &   5.2079414  &   6.9044853  \\
$z_1/M$    &  -7.2642136  &  -10.415883  &  -13.808971  \\
$p^x_1/M$  &  -0.0613136  &  -0.0483577  &  -0.0406335  \\
$p^z_1/M$  &   0.0334305  &   0.0279183  &   0.0242469  \\
$m^p_1/M$  &   0.3239234  &   0.3273810  &   0.3290810  \\
$m^H_1/M$  &   0.3356291  &   0.3355032  &   0.3351917  \\
$y_2/M$    &  -3.6321068  &  -5.2079414  &  -6.9044853  \\
$z_2/M$    &  -7.2642136  &  -10.415883  &  -13.808971  \\
$p^x_2/M$  &  -0.0613136  &   0.0483577  &   0.0406335  \\
$p^z_2/M$  &   0.0334305  &   0.0279183  &   0.0242469  \\
$m^p_2/M$  &   0.3239234  &   0.3273810  &   0.3290810  \\
$m^H_2/M$  &   0.3356566  &   0.3355216  &   0.3352024  \\
$z_3/M$    &  14.5284272  &   20.831766  &   27.617941  \\
$p^z_3/M$  &  -0.0668610  &  -0.0558366  &  -0.0484938  \\
$m^p_3/M$  &   0.3247293  &   0.3273813  &   0.3288641  \\
$m^H_3/M$  &   0.3313404  &   0.3320131  &   0.3323634  \\
$M\Omega$  &   0.0375000  &   0.0225000  &   0.0150000  \\
$D/M$      &   7.2642136  &   10.415883  &   13.808971  \\
\end{tabular}
\end{ruledtabular}
\label{table:ID-1}
\end{table}
\begin{table}
\caption{continuation of Table~\ref{table:ID-1}}
\begin{ruledtabular}
\begin{tabular}{lccc}
Config     &  3BH209      &  3BH211      &  3BH303  \\
\hline
$y_1/M$    &   8.4201027  &   11.117239  &   3.6321068  \\
$z_1/M$    &  -16.840205  &  -22.234477  &  -7.2642136  \\
$p^x_1/M$  &  -0.0361227  &  -0.0307979  &  -0.0613136  \\
$p^z_1/M$  &   0.0219565  &   0.0191084  &   0.0668610  \\
$m^p_1/M$  &   0.3299485  &   0.3308518  &   0.3171133  \\
$m^H_1/M$  &   0.3349484  &   0.3346311  &   0.3299497  \\
$y_2/M$    &  -8.4201027  &  -11.117239  &  -3.6321068  \\
$z_2/M$    &  -16.840205  &  -22.234477  &  -7.2642136  \\
$p^x_2/M$  &   0.0361227  &  -0.0307979  &   0.0613136  \\
$p^z_2/M$  &   0.0219565  &   0.0191084  &   0.0668610  \\
$m^p_2/M$  &   0.3299485  &   0.3308518  &   0.3171133  \\
$m^H_2/M$  &   0.3349593  &   0.3346393  &   0.3299874  \\
$z_3/M$    &   33.680411  &   44.468955  &   14.528427  \\
$p^z_3/M$  &  -0.0439130  &  -0.0382167  &  -0.1337220  \\
$m^p_3/M$  &   0.3296776  &   0.3305720  &   0.2955146  \\
$m^H_3/M$  &   0.3255030  &   0.3327501  &   0.3075078  \\
$M\Omega$  &   0.0112500  &   0.0075000  &   0.0375000  \\
$D/M$      &   16.840205  &   22.234477  &   7.2642136  \\
\end{tabular}
\end{ruledtabular}
\label{table:ID-2}
\end{table}
\begin{table}
\caption{continuation of Table~\ref{table:ID-1}}
\begin{ruledtabular}
\begin{tabular}{lccc}
Config     &  3BH307      &  3BH407      &  3BH507      \\
\hline
$y_1/M$    &   6.9044853  &   6.9044853  &   6.9044853  \\
$z_1/M$    &  -13.808971  &  -13.808971  &  -13.808971  \\
$p^x_1/M$  &  -0.0406335  &  -0.0406335  &  -0.0406335  \\
$p^z_1/M$  &   0.0484938  &   0.0727407  &   0.0969876  \\
$m^p_1/M$  &   0.3256512  &   0.3197642  &   0.3111199  \\
$m^H_1/M$  &   0.3323764  &   0.3275486  &   0.3206201  \\
$y_2/M$    &  -6.9044853  &  -6.9044853  &  -6.9044853  \\
$z_2/M$    &  -13.808971  &  -13.808971  &  -13.808971  \\
$p^x_2/M$  &   0.0406335  &   0.0406335  &   0.0406335  \\
$p^z_2/M$  &   0.0484938  &   0.0727407  &   0.0969876  \\
$m^p_2/M$  &   0.3256512  &   0.3197642  &   0.3111199  \\
$m^H_2/M$  &   0.3324005  &   0.3275903  &   0.3205268  \\
$z_3/M$    &   27.617941  &   27.617941  &  27.6179413  \\
$p^z_3/M$  &  -0.0969876  &  -0.1454815  &  -0.1939753  \\
$m^p_3/M$  &   0.3146486  &   0.2872890  &   0.2319451  \\
$m^H_3/M$  &   0.3208728  &   0.2987465  &   0.2539505  \\
$M\Omega$  &   0.0150000  &   0.0150000  &   0.0150000  \\
$D/M$      &   13.808971  &   13.808971  &   13.808971  \\
\end{tabular}
\end{ruledtabular}
\label{table:ID-3}
\end{table}
\begin{table}
\caption{Initial data parameters for the Burrau problem
(3BHTR1 and 3BHTR2),
as well as an off-center interaction of a binary with
a third black hole falling toward the binary near the $z$-axis (3BHOC).}
\begin{ruledtabular}
\begin{tabular}{lccc}
Config     &  3BHOC       &  3BHTR1      &  3BHTR2      \\
\hline
$x_1/M$    &   5.8677220  &   5.0000000  &   2.5000000  \\
$y_1/M$    &   4.0142579  &   15.000000  &   7.5000000  \\
$z_1/M$    &  -15.057925  &   0.0000000  &   0.0000000  \\
$p^x_1/M$  &  -0.0206448  &   0.0000000  &   0.0000000  \\
$p^y_1/M$  &   0.0297924  &   0.0000000  &   0.0000000  \\
$p^z_1/M$  &   0.0232357  &   0.0000000  &   0.0000000  \\
$m^p_1/M$  &   0.3298206  &   0.3000000  &   0.3000000  \\
$m^H_1/M$  &   0.3355713  &   0.3061460  &   0.3122950  \\
$x_2/M$    &  -5.8750250  &  -10.000000  &  -5.0000000  \\
$y_2/M$    &  -4.0215595  &  -5.0000000  &  -2.5000000  \\
$z_2/M$    &  -15.057934  &   0.0000000  &   0.0000000  \\
$p^x_2/M$  &   0.0206502  &   0.0000000  &   0.0000000  \\
$p^y_2/M$  &  -0.0297870  &   0.0000000  &   0.0000000  \\
$p^z_2/M$  &   0.0232348  &   0.0000000  &   0.0000000  \\
$m^p_2/M$  &   0.3298209  &   0.4000000  &   0.4000000  \\
$m^H_2/M$  &   0.3355903  &   0.4090649  &   0.4181309  \\
$x_3/M$    &   0.0073029  &   5.0000000  &   2.5000000  \\
$y_3/M$    &   0.0073015  &  -5.0000000  &  -2.5000000  \\
$z_3/M$    &   30.115859  &   0.0000000  &   0.0000000  \\
$p^x_3/M$  &  -5.3366e-6  &   0.0000000  &   0.0000000  \\
$p^y_3/M$  &  -5.3522e-6  &   0.0000000  &   0.0000000  \\
$p^z_3/M$  &  -0.0464705  &   0.0000000  &   0.0000000  \\
$m^p_3/M$  &   0.3292338  &   0.5000000  &   0.5000000  \\
$m^H_3/M$  &   0.3324495  &   0.5104157  &   0.5208322  \\
$M\Omega$  &   0.0150000  &   *********  &   *********  \\
$D/M$      &   14.223576  &   *********  &   *********  \\
\end{tabular}
\end{ruledtabular}
\label{table:ID-4}
\end{table}

\subsection{Binary---third-black-hole interactions} 
\label{sec:binary_plus_3}
Initial data families of quasi-circular black-hole binaries
in a hierarchical 3-body
system with a third companion relatively far away from the binary
were studied in Ref.~\cite{Campanelli:2005kr}.  That study was
based on the second-order post-Newtonian (2PN) approximation to the
3-body Hamiltonian (We provide the 3-body Hamiltonian for our
configurations in Appendix~\ref{ap:3-body-H}.). In this work we use the
2-body Hamiltonian to generate quasi-circular black-hole binary
configurations and then add a third black hole. This setup models
an `adiabatic' interaction of a binary with a third-black hole, where
the infall timescale is significantly faster than the timescale
for the binary to re-circularize through the emission of gravitational
radiation. Thus the binary starts off with the orbital parameters 
it would have if the third body were not present.

We evolve three sets of configurations with a third black hole falling
towards the center of a binary perpendicular to the binary's orbital
plane (which we take to be the $xy$ plane).  In all sets the third
 black hole is initially at a separation
equal to three times the binary's separation and all black holes contribute
equally to the total ADM mass given by Eq.~(\ref{eq:approx_MADM})
(As discussed below, they do not necessarily have similar horizon
masses.).
 In the first set we choose the third black hole
to have the momentum of a particle falling into the binary from
infinity (with zero speed). We obtain this momentum by assuming that
the binary separation is fixed and then treat the resulting effective
two-body problem via Newtonian mechanics. We find that the third black
hole has momentum
\begin{equation}
p_3^z = -P_{0} = -\frac{4}{9} M \sqrt{\frac{M}{D \sqrt{37}}},
\label{eq:P0}
\end{equation}
where $D$ is the initial binary separation and $M$ is the total mass
(the binary has equal momentum in the opposite direction).
In addition, we evolve configurations with $p_3^z = -2 P_{0}$, 
$p_3^z = -3 P_{0}$, and $p_3^z = -4 P_{0}$. We denote these configurations
by 3BHYXX, where $p_3^z = -({\rm Y}-1) P_{0}$ and
XX indicates the binary's initial separation (XX is {\em not}  equal
to the binary separation).
Initial data parameters for the configurations are given in
Tables~\ref{table:ID-1}--\ref{table:ID-3}. In all cases the members of
the binary are given linear momenta consistent with 3PN quasi-circular
orbits when ignoring the contribution of the third black hole. We also
evolve identical configurations of mass, initial positions, and
initial momenta using a Newtonian code to compare Newtonian dynamics
with the results from the fully nonlinear general relativistic
evolutions.
Note that the horizon mass of the third black hole is not equal to the
horizon mass of the other two (the differences in the horizons masses
between the members of the binary is a due solely to finite-difference errors
in the isolated horizon algorithm~\cite{Dreyer:2002mx}). This difference in horizon mass
becomes smaller as the binary separation is increased and larger as
the third-black-hole momentum is increased. It would be interesting to
explore the sequence where all three black holes have the same horizon
mass to see if qualitatively different results are obtained.

\begin{figure}
\begin{center}
\caption{The $xy$ projection of the binary-separation trajectory when the third
black hole falls towards the center of the binary along the $z$ axis 
with initial momentum $p_3^z = -P_{0}$. The coordinates have been
rescaled by the initial binary separation. The solid curves are the
GR trajectories, while the dot-dashed curves are the Newtonian
trajectories. $N_{\rm limit}$ is the Newtonian trajectory
for $D\to\infty$ (i.e.\ $\Omega\to0$). There seems to be a critical
separation between 3BH207 and 3BH209 where the system transitions from
a prompt triple merger to an elliptical binary plus third black hole
(we indicate which systems triple
merge with a $*$ in the legend).
}
\includegraphics[width=3.3in]{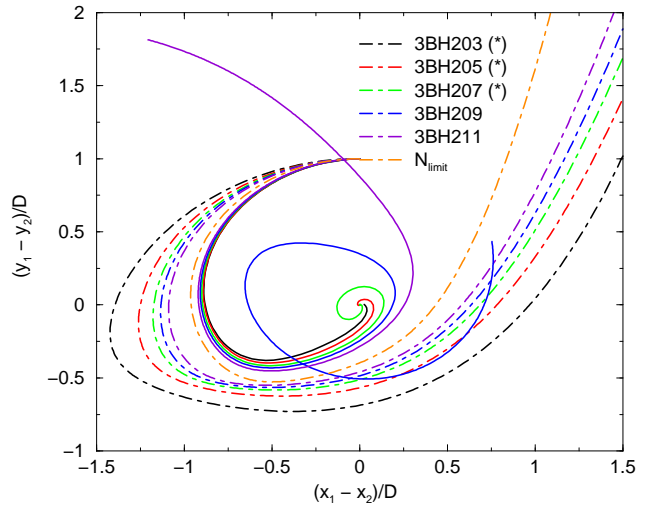}
\label{fig:free_throw_xy_difftrack}
\end{center}
\end{figure}

\begin{figure}
\begin{center}
\caption{The $z$-component (rescaled by the initial binary
separation) of the trajectory of the third black hole
with initial momentum $p_3^z = -P_{0}$. $\Omega$ is the
binary's initial frequency. The solid curves are the
GR trajectories, while the dot-dashed curves are the Newtonian
trajectories. $N_{\rm limit}$ is the Newtonian trajectory
for $D\to\infty$ (i.e.\ $\Omega\to0$). Systems that triple-merge
are indicated by a $*$ in the legend.
}
\includegraphics[width=2.7in]{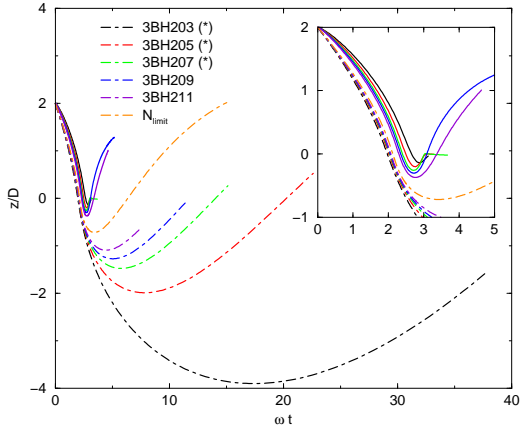}
\label{fig:free_throw_z_track}
\end{center}
\end{figure}

In Fig.~\ref{fig:free_throw_xy_difftrack} we show the binary
separation vector $\vec r  = \vec r_1 - \vec r_2$, projected onto the
$xy$ plane, for the first set of configurations with $p_3^z = -P_{0}$, as well as the Newtonian
results for these configurations (the initial third-black-hole to
binary separation is three times the binary's initial separation).
After rescaling
by the binary's initial separation, we see that, initially, all
binaries perform similar (distorted) elliptical trajectories. However,
the configurations with close binary separations show a prompt triple
merger of the binary and third black hole (we indicate which systems triple
merge with a $*$ in the legend). As the binary's separation
is increased, it undergoes more of an orbit before merging, and
even receives a substantial kick from the third black hole. The plot
seems to indicate that there is a critical value~\cite{Pretorius:2007jn}
$D_{\rm crit}$,
where the binary quickly merges for $D < D_{\rm crit}$ and is given a
substantial kick above this value. The Newtonian trajectories show the
opposite trend, where the larger the initial binary separation, the
more compact the orbit. The two systems appear to be converging to the
same orbit as $\Omega\to0$. The Newtonian orbits are scale invariant when
the orbital momentum corresponds to Newtonian circular trajectories (for the
two particles in the binary). As $\Omega$ is decreased and the binary
separation gets large, the difference between the post-Newtonian and
Newtonian momenta becomes negligible and the system becomes scale
invariant. To explicitly show this, we
evolved similar configurations, but with $M \Omega = 10^{-5}$ and
$M \Omega=10^{-10}$, and confirmed that the Newtonian orbit approaches
the limiting trajectory given in the figure.  In
Fig.~\ref{fig:free_throw_z_track} we show the position of the third
black hole versus time. Note here as well that the Newtonian and GR
trends show the opposite behaviors. In the Newtonian case, the smaller
$D$ is (i.e.\ the larger $\Omega$ is) the more likely the third
particle is to pass through the binary, while in the GR simulations
the third black hole always merges with the binary for small $D$. 
The GR and Newtonian
evolutions approach each other as $D$ tends toward infinity. 

\begin{figure}
\begin{center}
\caption{
The $xy$ projection of the binary-separation trajectory when the third
black hole falls towards the center of the binary along the $z$ axis 
with initial momentum $p_3^z = -P_{0}, -2 P_{0}, -3 P_{0}, -4 P_{0}$. The coordinates have been
rescaled by the initial binary separation. The solid curves are the
GR trajectories, while the dot-dashed curves are the Newtonian
trajectories. Note that 3BH207 and 3BH507 result in bound binaries in
Newtonian theory. There appears to be a critical momentum where
the system transitions from a prompt merger to a highly-elliptical,
or perhaps even hyperbolic, orbit. Systems that triple-merge are indicated
by a $*$ in the legend.
}
\includegraphics[width=3.3in]{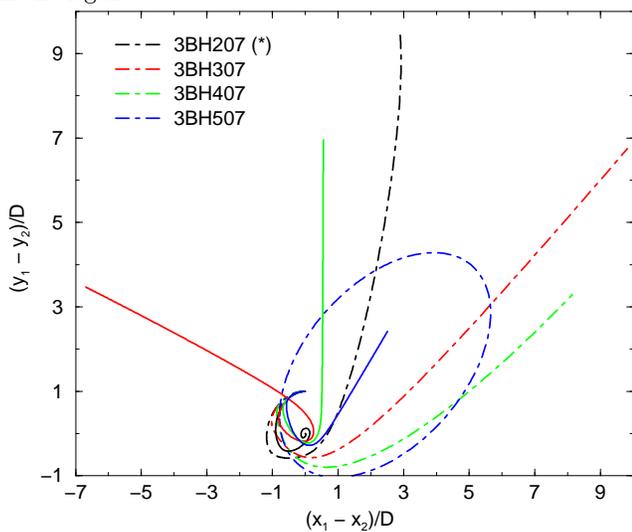}
\label{fig:forced_throw_xy_track}
\end{center}
\end{figure}

\begin{figure}
\begin{center}
\caption{
The $z$-component (rescaled by the binary initial
separation) of the trajectory of the third black hole
with initial momentum $p_3^z = -P_{0}, -2 P_{0}, -3 P_{0}, -4 P_{0}$. $\Omega$ is the
binary's initial frequency. The solid curves are the
GR trajectories, while the dot-dashed curves are the Newtonian
trajectories. Systems that triple-merge are indicated
by a $*$ in the legend.}
\includegraphics[width=2.7in]{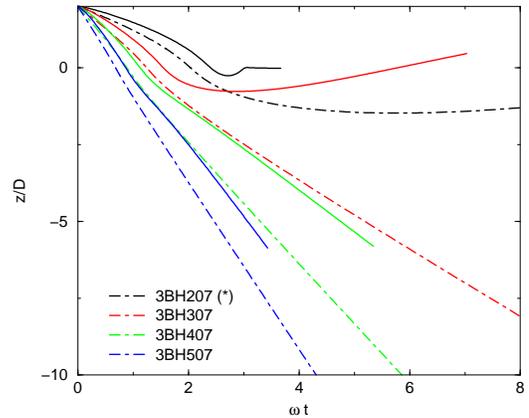}
\label{fig:forced_throw_z_track}
\end{center}
\end{figure}

In Figs.~\ref{fig:forced_throw_xy_track}~and~\ref{fig:forced_throw_z_track}
we explore how the three-body system behaves when the initial $z$-momentum
is increased, while still keeping the initial third-black-hole to binary
 separation at three times the binary's initial separation. 
We choose a relatively large initial binary separation of
$D = 13.808971 M = 20.7134565 M_{\rm B}$ ($M_{\rm B}$ is the binary's
mass). In the absence of the third
black hole, this binary would complete approximately $25$ orbits
before merging. In Newtonian theory, configurations 3BH207, 3BH407, and
3BH507 (i.e.\ $p_3^z = -P_{0}, -3 P_{0}, -4 P_{0}$) result
in a bound binary with the third particle ejected to infinity, while
for configuration 3BH307 the entire system is disrupted; leading to three
free particles. Note also that in Newtonian theory the third black hole
passes through the binary and is ejected toward $z=-\infty$ for all
configuration except 3BH207. The GR simulations on the other hand seem
to indicate that the binary is disrupted for all configuration except
2BH207 (which triple merges). Interestingly the behavior of the third particle
approaches the Newtonian behavior as $p_3^z$ is increased (as can be seen
in Fig.~\ref{fig:forced_throw_z_track}). Note that, in the GR simulation of
3BH307, the third black
hole was not able to escape to $z=-\infty$, but rather was bounced toward
$z=+\infty$. 

\begin{figure}
\begin{center}
\caption{The $xy$ projection of the binary-separation trajectory when the third
black hole falls towards the center of the binary along the $z$ axis
with initial momentum $p_3^z = -P_{0}, -2 P_{0}$ at two different
initial binary separations. The coordinates have been
rescaled by the initial binary separation. Here the critical separation
between prompt-merger and (possible) disruption is a function
of both the binary separation and the initial $z$-momentum. Systems that triple-merge are indicated
by a $*$ in the legend.
}
\includegraphics[width=2.7in]{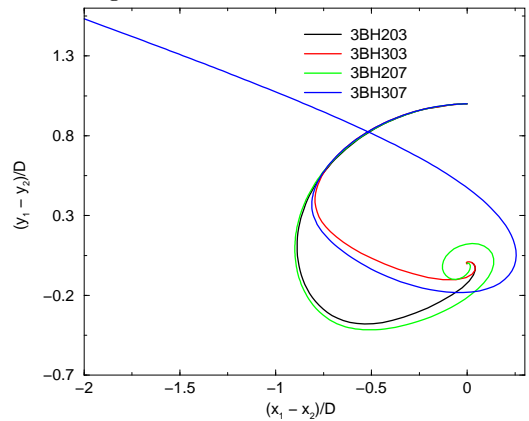}
\label{fig:forced_throw_compare_sep}
\end{center}
\end{figure}

We conclude our analysis of this three-black-hole configuration by comparing
the behavior of the system when the $z$-momentum is doubled and the binary
separation is increased (while still keeping the initial third-black-hole
to binary separation at three times the initial binary separation).
 In Fig.~\ref{fig:forced_throw_compare_sep} we show
the $xy$ trajectories for configurations 3BH203, 3BH303, 3BH207, 3BH307.
Note that for the closer binary (3BH203, 3BH303) increasing the momentum
of the third black hole (which increases the eccentricity of the binary)
causes the binary to merge sooner, while for the larger binary this same
increase leads to a (possible) disruption of the binary.

\begin{figure}
\begin{center}
\caption{The $xy$ projection of the Newtonian trajectories
for 3BHOC. The initial positions of the three particles are indicated
by filled circles. Note that the initial P1--P2 binary is disrupted and
a new P1--P3 binary is formed (The new P1--P3 binary is
inclined with respect to the $xy$ plane.). The initial motion of P3 is
essentially along the $z$-axis and is thus not apparent.}
\includegraphics[width=2.7in]{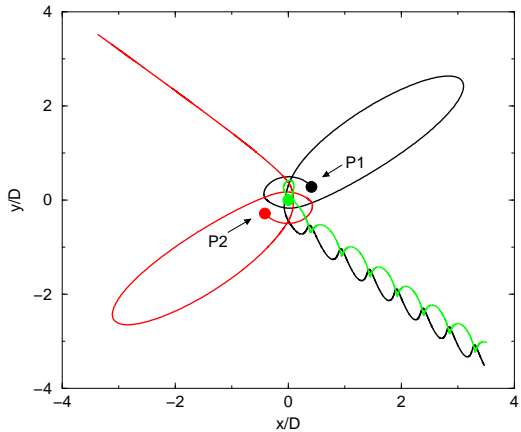}
\label{fig:config_OC_xy_newton}
\end{center}
\end{figure}

\begin{figure}
\begin{center}
\caption{The $xy$ projection of the binary-separation trajectory for
3BHOC, as well as 3BH207 and the Newtonian trajectory.
Note that 3BHOC is more elliptical (and thus merges sooner) than
3BH207.}
\includegraphics[width=2.7in]{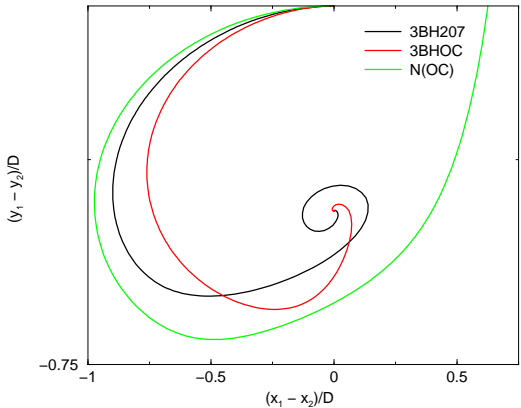}
\label{fig:compare_207_oc_xy}
\end{center}
\end{figure}

It is interesting to compare the dynamics of this three-black-hole
system under modified initial conditions that reduce the
symmetry of the problem. We began this type of analysis by considering
a purely Newtonian system consisting of a circular binary with
orbital frequency $M \Omega = 0.015$ and mass $M_{B} = 2/3 M$ (similar
to 3BHY07, but with Newtonian, rather than post-Newtonian, orbital
momenta) and a third particle of mass $1/3 M$ located at $(x,y,z)/D = 
(0.0099, 0.0099, 1000)$. We then evolved this configuration using purely
Newtonian evolution until the third-particle---binary separation was
$\sim 3 D$. We then took the Newtonian position and momentum parameters
and evolved them using full GR. In the purely Newtonian evolution, this 3-body
system leads to an exchange of partners in the binary, where particle 2
is ejected from the system and particles 1 and 3 form a new binary 
with eccentricity $e = 0.645132$ (See
Fig.~\ref{fig:config_OC_xy_newton}).
The full GR simulations, on the other hand, show a triple merger that
merges faster than 3BH207 (which likely due to our using Newtonian,
rather than post-Newtonian, orbital momenta). In Fig.~\ref{fig:compare_207_oc_xy}
we show the $xy$ projection of the initial binary trajectory
for 3BHOC, as well as 3BH207 and the Newtonian trajectory.

\subsection{Burrau Three-Body Configuration}
\begin{figure}
\begin{center}
\caption{The full Newtonian tracks for the 3BHTR1
configurations (3BHTR2 is obtained by  rescaling by a factor of 1/2).
Note that particle 1 is ejected and moves towards the
upper right of the figure, while particles 2 and 3 form an elliptical
binary that moves towards the lower left. Initial positions are indicated by
a filled circles. Particle 1 (mass $0.3M$) is the one on top,
particle 2 (mass $0.4M$) is the lower left, and particle 3 (mass $0.5M$) is
the lower right. The arrows indicate the trajectories of the recoiled binary
and lone particle.}
\includegraphics[width=2.7in]{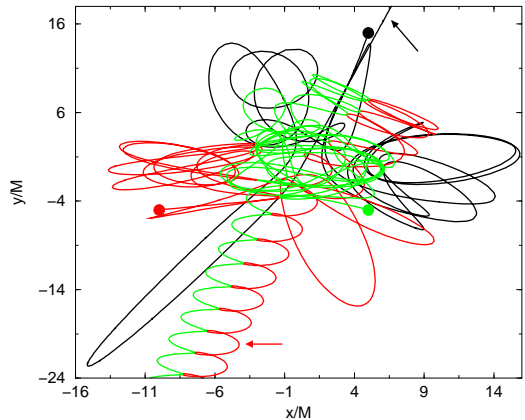}
\label{fig:tr_newton_xy_track}
\end{center}
\end{figure}

\begin{figure}
\begin{center}
\caption{The GR and Newtonian tracks for configurations
3BHTR1 and 3BHTR2, the latter rescaled by a factor of 2 
(the Newtonian trajectories for these configurations
are identical up to a scaling). After rescaling, the two GR
trajectories are very similar (see right inset), and differ significantly
from the Newtonian trajectories after the first BH2---BH3 close
encounter (see left inset), which leads to a merger in GR. }
\includegraphics[width=2.7in]{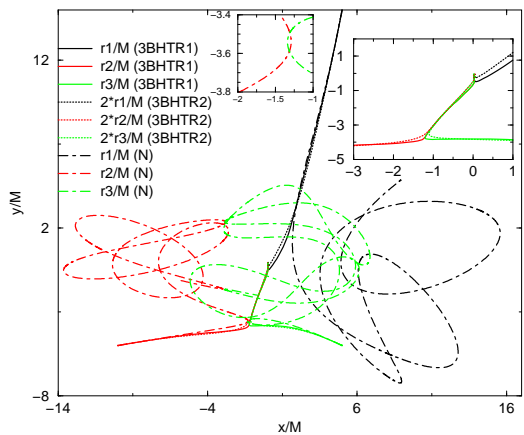}
\label{fig:tr1_tr2_xy_track}
\end{center}
\end{figure}

Configurations 3BHTR1 and 3BHTR2 belong to a set of configurations
known as the Burrau three-body
problem~\cite{Burrau1913,Szebehely1967,Valtonen1995}.  This system
consists of three particle, initially at rest, arranged at the vertices of a right
triangle with sides of length $3 \rho$, $4 \rho$, $5 \rho$ and masses
$3 \mu$, $4 \mu$, $5 \mu$, where the particle of mass $i\,\mu$ is
located on the vertex opposite the side of length $i\,\rho$. After
suitable rescaling, the Newtonian trajectories are independent of
$\rho$ and $\mu$. In Newtonian theory, this configuration will lead to
particles 2 and 3 forming a highly elliptical binary (with ellipticity
$e\approx0.989$), and particle 1 and the 2--3 binary ejected in
opposite directions (See
Fig.~\ref{fig:tr_newton_xy_track}). 
The trajectories of the corresponding black holes in a full GR
simulation are quite different (although, if $\mu$ is kept fixed and
$\rho$ is increases, eventually the GR simulations will reproduce the
Newtonian trajectories). Here we see that the trajectories scale
reasonably well with $\rho$ (i.e.\ compare 3BHTR1 and 3BHTR2 in 
Fig.~\ref{fig:tr1_tr2_xy_track}), but rather than forming a binary and
free particle, this system quickly merges to form a single black hole.

The Newtonian, and even 2PN, trajectories for this three-black-hole
problem will not agree with the GR trajectories if the closest
approach of any two BHs is within about ten times the combined
mass of the two BHs. To analyze the behavior of the system at different
scales, we set the masses of the three holes
to $3m$, $4m$, $5m$, and set the initial separations to
$r_{13} = 200 m \rho$, $r_{23}=150 m \rho$, $r_{12} = 250 m \rho$.
The initial merger of BH2 and BH3 seen in the GR simulation corresponds to
a close approach of $r_{23}/(m_2 + m_3) = 0.054 \rho$. 
Thus,
for the initial close encounter not to lead a quick merger, we need
$\rho\sim185$, which means that the initial value of $r_{23}$ will
need to be larger than $r_{23} = 1.4\times10^{-9} pc (m/M_{\odot})$. However,
the closest approach over the entire trajectory is
 $r_{12}/(m_2 + m_3) =0.0023\rho$, and in order for this approach not
to lead to a prompt merger the initial value of $r_{23}$ would
need to be $r_{23} = 3.3\times10^{-8}pc (m/M_{\odot})$.

\begin{figure}
\begin{center}
\caption{The $(\ell=2,m=2)$ mode of $\psi_4$ for 3BHTR1 showing both
the two mergers. Note that the
imaginary part of this mode is more sensitive to the second merger.
The two mergers are clearly seen in the absolute value of $\psi_4$. }
\includegraphics[width=2.7in]{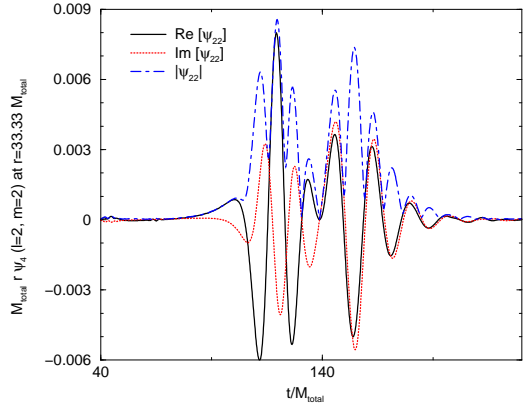}
\label{fig:tr1_wave}
\end{center}
\end{figure}

\begin{figure}
\begin{center}
\caption{The $(\ell=2,m=2)$ mode of $\psi_4$ for 3BHTR2. Here the real
part of $\psi_4$ seems to only indicate the presence of one merger,
while the second merger is more clearly visible in the imaginary part.}
\includegraphics[width=2.7in]{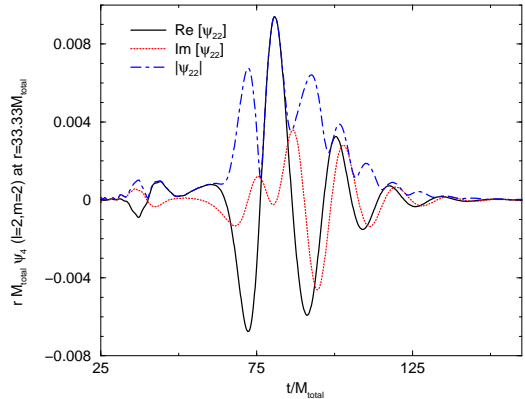}
\label{fig:tr2_wave}
\end{center}
\end{figure}

The waveforms from 3BHTR1 and 3BHTR2 show double mergers, with the two
mergers more closely spaced in 3BHTR2. In
Figs.~\ref{fig:tr1_wave}~and~\ref{fig:tr2_wave} we show the
$(\ell=2,m=2)$ mode of $\psi_4$ extracted at $r=40M =
33.33M_{ADM}$ (here $M_{ADM} = 1.2 M$).
Interestingly, the real part of the $(\ell=2,m=2)$ mode is more
sensitive to the initial merger (of BHs 2 and 3), while the imaginary
part is more sensitive to the second merger (of the 2--3 remnant with
BH 1).
The radiated energy and angular momentum for configuration 3BHTR1
were $E_{\rm rad}/M_{\rm ADM} = (8.5\pm1.0)\times10^{-4}$ and
 $J_{\rm rad}/M_{\rm ADM}^2 = (2.8\pm2.1)\times10^{-4}$,
while for configuration 3BHTR2 the radiated quantities were
$E_{\rm rad}/M_{\rm ADM} = (6.6\pm0.2)\times10^{-4}$ and
$J_{\rm rad}/M_{\rm ADM}^2 = (1.2\pm0.6)\times10^{-4}$. The recoil
velocities for 3BHTR1 and 3BHTR2 were $(4.1\pm2.0)\KMS$ and
$(3.0\pm0.6)\KMS$ respectively.

\subsection{Eighth-Order Accuracy}
\begin{figure}
\begin{center}
\caption{The $(\ell=2,m=2)$ mode of $\psi_4$ for Q38 using both
4th and 8th-order algorithms. Note that the phase error in the $M/80$
8th-order waveform is apparently smaller than the phase error in the $M/120$
4th-order waveform.}
\includegraphics[width=2.7in]{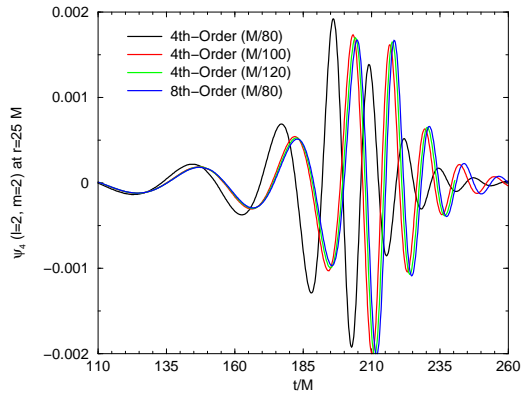}
\label{fig:q38_8th_4th}
\end{center}
\end{figure}

\begin{figure}
\begin{center}
\caption{The $z$ component of the trajectory of BH1 for configuration 3BH102. This
component should be zero by symmetry. The 8th-order algorithm
produces an error that is $3.33$ times smaller. Note that the error in
the 4th-order trajectory is smaller than the central resolution of $M/80$ until
$t\sim 290$.}
\includegraphics[width=2.7in]{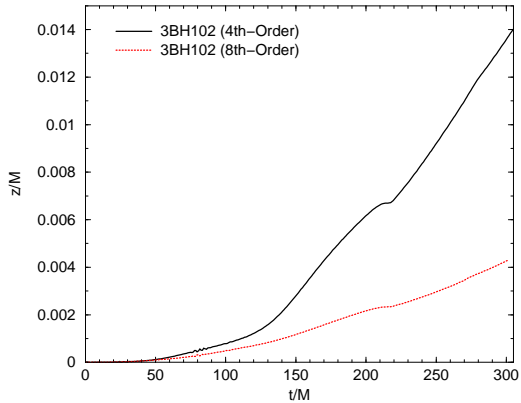}
\label{fig:3BH102_4th_8th_compare}
\end{center}
\end{figure}

\begin{figure}
\begin{center}
\caption{The $xy$ projection of the trajectory of all three BHs in 3BH102. Solid
curves were produced using the fourth-order algorithm; dot-dashed curved using
the eighth-order algorithm.
}
\includegraphics[width=2.7in]{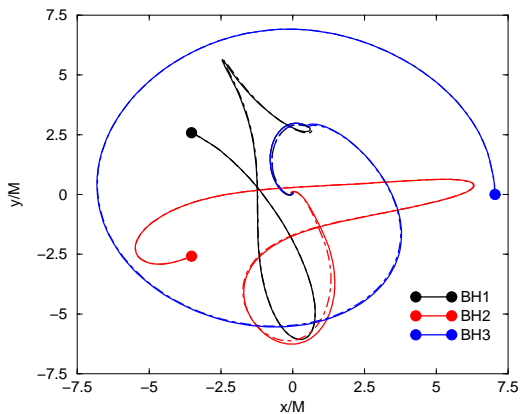}
\label{fig:3BH102_4th_8th_compare_xy}
\end{center}
\end{figure}

\begin{figure}
\begin{center}
\caption{The real part of the $(\ell=2,m=2)$ mode of $\psi_4$ for
configuration 3BH1 of Ref.~\cite{Campanelli:2007ea}
with central resolutions of $M/80$, $M/96$, and
$M/115.2$, along with a convergence plot of data. Note the
fourth-order convergence, and the smallness of the waveform amplitude.}
\includegraphics[width=2.7in]{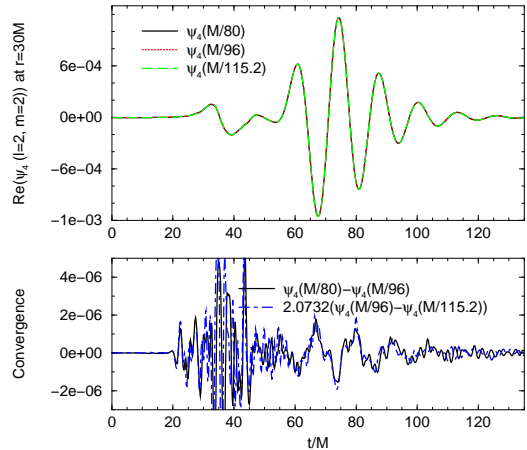}
\label{fig:cf1_conv}
\end{center}
\end{figure}

We evolved several configurations, including Q38 of
 Refs.~\cite{Krishnan:2007pu,Lousto:2007db} (which is a quasi-circular,
non-spinning binary with mass ratio 3:8) and 
3BH102 of Ref.~\cite{Campanelli:2007ea} (which is
a three-black-hole configuration with planar orbits),
 using our standard fourth-order code and the new eighth-order
code. In all cases we used the same grid structure with a maximum
resolution of $M/80$. In Fig.~\ref{fig:q38_8th_4th} we plot the $(\ell=2,
m=2)$ component of $\psi_4$ for the Q38 configuration with a 
Gamma-driver parameter of $\eta=2/M$. As pointed out in~\cite{Lousto:2007db}, this choice
of $\eta$ leads to a low effective resolution for our lowest resolution
run (i.e. $M/80$), leading to a large phase error. The corresponding 
eighth-order run appears to be more accurate than the $M/120$ fourth-order
run (i.e.\ its phase is closer to that predicted by extrapolating using
the fourth-order $M/100$ and $M/120$ runs).
Eighth-order also significantly reduces the error in three-black-hole
 simulations. To test this, we evolved configuration 3BH102 (which
consists of three equal-mass black holes, labeled BH1 -- BH3 in
Fig.~\ref{fig:3BH102_4th_8th_compare_xy}, in a
planar orbit) using a
shifted grid structure (i.e.\ one not centered on the origin) that does
not exhibit the $z$-reflection symmetry of this
configuration. In Fig.~\ref{fig:3BH102_4th_8th_compare} we plot the
$z$-component of the trajectory for BH1 (which should be identically
zero). Note that late-time error in $z$ is reduced by a factor of
$3.33$ when going from 4th to 8th-order for this coarse grid structure,
which may be crucial for longer term evolutions.
The $xy$ trajectories do not change qualitatively for 3BH102 when moving
to eighth-order, as is apparent in Fig.~\ref{fig:3BH102_4th_8th_compare_xy}.
We will present results from convergence and efficiency tests in a
forthcoming paper. Here we note that we found fourth-order convergence (See
Fig.~\ref{fig:cf1_conv}) in the waveform for three-black-hole systems
using our standard algorithm.

\section{Discussion}\label{sec:disc}
Our fully nonlinear evolutions of three-black-hole systems demonstrate
dissipative General Relativistic effects due to emission of
gravitational radiation and black-hole mergers, which do not have
Newtonian (or even 2PN) counterparts, that dramatically change the
qualitative behavior of the system. These effects become important
when two objects approach each other closer than about 100 times their
combined mass~\cite{Lousto:2007ji} 
(or $d\sim 100 G (m_1+m_2)/c^2$ in more conventional
units) and become dominant when the two objects approach within about
10 times their combined mass. Even when the effect is small, the
sensitivity of three (and more) body encounters to small perturbations
makes including PN and GR effects important for obtaining the correct
dynamics of an N-body simulation. The GR effects in particular
provide a natural regularization of the Newtonian problem on small
scales.

An important problem in galactic dynamics is determining how the
merger process of the two (initially) central black holes proceeds
after the collision of the host galaxies~\cite{GouldRix:2000,
Milosavljevic:2001vi,Yu:2002}.  In our
simulations we confirm that the resulting binary will, in general,
become highly elliptical if it interacts with a similar mass black
hole. These eccentric orbits will drive the binary merger due to
emission of gravitational radiation. However, we also found that 
the interaction can produce an immediate merger of the triple system.
If we set physical scale of our simulation by setting the black-hole
masses to  $10^9M_\odot$, then the prompt mergers that we observe
occur at separations of the order of
milli-parsecs. 
These three-body interactions provides a possible mechanism for resolving
the ``last parsec problem.''~\cite{Milosavljevic:2002ht}
We will present results from the effects of unequal-masses, spins,
and further separated binaries in a forthcoming paper, where we 
will also explore the potential for critical phenomena.
It is important to note that these three-black-hole interactions
provide a mechanism for producing highly-elliptical close-binaries (which
would otherwise have circularized due to emission of gravitational
radiation during the inspiral) like 
those studied in Ref.~\cite{Hinder:2007qu}.

\acknowledgments
We thank Manuela Campanelli, Frans Pretorius, David Merritt, Alessia Gualandris,
 and Hiroyki Nakano for valuable discussions.
We gratefully acknowledge NSF for financial support
from grant PHY-0722315,  PHY-0701566, PHY 0714388, and PHY 0722703.
We also thank Hans-Peter Bischof for producing  3-D 
visualizations of the three-black-hole configurations introduced in this paper.
Computational resources were provided by the NewHorizons cluster at RIT and
the Lonestar cluster at TACC.

\appendix
\section{2PN orbits}
\label{ap:3-body-H}
In this paper we did not include all Newtonian or Post-Newtonian
3-body interactions when producing initial data for quasi-circular orbits.
However, these interaction can be taken into account, up to 2PN, using
the 
Hamiltonian of the 3-body interactions in the
ADM gauge as given by Eq.~(5) in Ref.~\cite{Schaefer87} (and 
Eqs.~(1),(2),(A1) in Ref.~\cite{Lousto:2007ji}).

Our configurations are characterized by
\begin{subequations}
\label{3BHID1}
\begin{eqnarray}
&&m_1=m_2=m_3=m\\
&&x_1=0, y_1=r/2, z_1=-r\\
&&x_2=0, y_2=-r/2, z_2=-r\\
&&x_3=0, y_3=0, z_3=2r\\
&&p^x_1=-J/r, p^y_1=0, p^z_1=-p^z_3/2\\
&&p^x_2=J/r, p^y_2=0, p^z_2=-p^z_3/2\\
&&p^x_3=0, p^y_3=0, \\
&&p^z_3=-\frac{4}{3}m\sqrt{\frac{3 m}{r \sqrt{37}}},
\end{eqnarray}
\end{subequations}
where $r$ is the separation of the binary,  the initial
third-body---binary separation is $3 r$, and the initial linear
momentum of the third body corresponds to the approximate
momentum it would have after falling toward the binary from
infinity.

The process for computing the quasi-circular orbits proceeds as
follows.  We first choose a value of the orbital angular momentum,
$J$, and  then compute the linear momenta of the holes as given above
in Eqs.~(\ref{3BHID1}).  We then compute the Hamiltonian,
\begin{widetext}
\begin{eqnarray}\label{HID1}
H = &&(3241792 \left(19+\sqrt{37}\right) J^6+21904
   r \left(4 \left(-1369+269 \sqrt{37}\right)
   m^3-37 \left(19+\sqrt{37}\right) \left(8 m^2-3
   {p_3^z}^2\right) r\right) J^4+ \nonumber \\
   && 148 r^2
   \left(1152 \left(90781+8503 \sqrt{37}\right)
   m^6-8 \left(296 \left(9583+709
   \sqrt{37}\right) m^2+\left(13801+19075
   \sqrt{37}\right) {p_3^z}^2\right) r
   m^3+ \right. \nonumber \\
   &&\left. 1369 \left(19+\sqrt{37}\right) \left(128
   m^4-16 {p_3^z}^2 m^2+3 {p_3^z}^4\right)
   r^2\right) J^2+r^3 \left(-128
   \left(9888649+1247443 \sqrt{37}\right)
   m^9+ \right. \nonumber \\
   &&\left.9472 \left(74 \left(1079+125
   \sqrt{37}\right) m^2+3 \left(100109+6140
   \sqrt{37}\right) {p_3^z}^2\right) r m^6-4
   \left(175232 \left(851+113 \sqrt{37}\right)
   m^4+ \right.\right. \nonumber \\
   &&\left.\left.592 \left(143449+85651 \sqrt{37}\right)
   {p_3^z}^2 m^2+3 \left(26603-429895
   \sqrt{37}\right) {p_3^z}^4\right) r^2
   m^3+\right.\nonumber \\
   &&\left.151959 \left(19+\sqrt{37}\right)
   {p_3^z}^2 \left(128 m^4-24 {p_3^z}^2
   m^2+11 {p_3^z}^4\right)
   r^3\right))/(25934336 \left(19+\sqrt{37}\right)
   m^5 r^6),
\end{eqnarray}
\end{widetext}
and find the value of $r$ that gives a local minimum (i.e.\ $\partial_rH=0$)~\cite{Campanelli:2005kr}, which gives the
value of the separation of the binary in a quasi-circular orbit.
Some of the results of this process are summarized in Table \ref{table:ID1}.

\begin{table}
\caption{2PN Initial data parameters for quasi-circular
orbits of binary black holes in the presence of a third black hole.
All black holes have mass $m = M/3$.}
\begin{ruledtabular}
\begin{tabular}{lll}
$J/M^2$           & $r/M$            & $p_{3}^z/M$\\
\hline
0.30 & 3.01288 & -0.103819\\
0.35 & 3.61132 & -0.0948274\\
0.40 & 4.58254 & -0.0841810\\
0.45 & 6.33147 & -0.0716168\\
0.50 & 8.85247 & -0.0605668\\
0.55 & 11.7723 & -0.0525215\\
0.60 & 14.9633 & -0.0465858\\
0.65 & 18.4081 & -0.0420013\\
0.70 & 22.1080 & -0.0383259\\
0.75 & 26.0665 & -0.0352960\\
0.80 & 30.2863 & -0.0327449\\
0.85 & 34.7695 & -0.0305610\\
0.90 & 39.5177 & -0.0286663\\
0.95 & 44.5321 & -0.0270042\\
1.00 & 49.8134 & -0.0255325\\
\end{tabular} \label{table:ID1} 
\end{ruledtabular} 
\end{table}

\bibliographystyle{apsrev}
\bibliography{../../../Lazarus/bibtex/references}

\end{document}